\documentstyle[12pt]{article}

\begin{document}

\author{C. Bizdadea\thanks{%
e-mail address: bizdadea@hotmail.com}, M. G. Mocioac\~a 
and S. O. Saliu\thanks{%
e-mail address: osaliu@central.ucv.ro} \\
Department of Physics, University of Craiova\\
13 A. I. Cuza Str., Craiova RO-1100, Romania}
\title{On the `Irreducible' Freedman-Townsend Vertex }
\maketitle

\begin{abstract}
An irreducible cohomological derivation of the Freedman-Townsend vertex in
four dimensions is given.

\end{abstract}

The problem of consistent interactions that can be introduced among gauge
fields in such a way to preserve the number of gauge symmetries \cite{alpha1}%
--\cite{alpha3} has been reformulated as a deformation problem of the master
equation \cite{def} in the framework of the antifield BRST formalism \cite{1}%
--\cite{5}. That deformation setting was applied to Chern-Simons models \cite
{def}, Yang-Mills theories \cite{alpha4} and two-form gauge fields \cite{23}%
. The deformation procedure for two-form gauge fields employs a reducible
BRST background.

The purpose of this letter is to reanalyze the problem of constructing
consistent interactions among two-form gauge fields (in four dimensions),
but following an {\it irreducible} BRST line in spite of the reducibility
present within the initial model. Our method contains two basic steps.
First, starting with abelian two-form gauge fields in first-order version
(obtained by means of adding some auxiliary fields), we construct an
irreducible BRST symmetry associated with the reducible one and show that we
can substitute the reducible symmetry by the irreducible one. Second, we
consistently deform the solution to the master equation associated with the
irreducible BRST symmetry. In this manner we obtain precisely the well-known
Freedman-Townsend vertex \cite{17}, the deformed gauge symmetries and also
the deformed solution to the master equation. However, our deformed solution
to the master equation differs from that obtained in the literature \cite{19}%
--\cite{22} by the fact that on the one hand our method introduces some
additional fields (necessary at the construction of the irreducible BRST
symmetry) which do not contribute to new type of couplings, and, on the
other hand, the deformed BRST transformations resulting from our formalism
do not involve the antifields due to the absence of terms quadratic in the
antifields in the deformed solution to the master equation, so the
gauge-fixed BRST symmetry is off-shell nilpotent. To our knowledge, such an
irreducible procedure for the Freedman-Townsend model has not previously
been published, our method establishing thus a new result.

The starting point is the Lagrangian action for abelian two-form gauge
fields in first-order form (also known as the abelian Freedman-Townsend
model) 
\begin{equation}
\label{1}S_0^L\left[ A_\mu ^a,B_a^{\mu \nu }\right] =\frac 12\int d^4x\left(
-F_{\mu \nu }^aB_a^{\mu \nu }+g_{ab}A_\mu ^aA^{b\mu }\right) , 
\end{equation}
where $B_a^{\mu \nu }$ is an antisymmetric tensor field, the field strength
of $A_\mu ^a$ is defined by $F_{\mu \nu }^a=\partial _{\left[ \mu \right.
}A_{\left. \nu \right] }^a$, with $\left[ \mu \nu \right] $ expressing
antisymmetry with the indices between brackets, and $g_{ab}$ is an
invertible, symmetric and constant matrix. It is simply to see that if we
eliminate the auxiliary fields $A_\mu ^a$ on their equations of motion, we
recover the action of free abelian two-form gauge fields. Action (\ref{1})
is invariant under the gauge transformations $\delta _\epsilon B_a^{\mu \nu
}=\varepsilon ^{\mu \nu \lambda \rho }\partial _\lambda \epsilon _{\rho a}$, 
$\delta _\epsilon A_\mu ^a=0$, where $\varepsilon ^{\mu \nu \lambda \rho }$
is the antisymmetric symbol in four dimensions. The above gauge
transformations are off-shell first-stage reducible as if we take $\epsilon
_{\rho a}=\partial _\rho \epsilon _a$, then $\delta _\epsilon B_a^{\mu \nu
}=0$.

The reducible Lagrangian BRST symmetry corresponding to the model described
by action (\ref{1}), $s_R=\delta _R+\gamma _R$, contains two main pieces,
namely, the Koszul-Tate differential $\delta _R$ and a model of longitudinal
derivative along the gauge orbits $\gamma _R$. In the case of our model, the
generators of the Koszul-Tate complex are the fermionic antighost number one
antifields $B_{\mu \nu }^{*a}$ and $A_a^{*\mu }$, the bosonic antighost
number two antifields $\eta ^{*a\mu }$ and the fermionic antighost number
three antifields $C^{*a}$. The definitions of $\delta _R$ read as 
\begin{equation}
\label{4}\delta _RB_a^{\mu \nu }=0,\;\delta _RA_\mu ^a=0, 
\end{equation}
\begin{equation}
\label{5}\delta _RB_{\mu \nu }^{*a}=\frac 12F_{\mu \nu }^a,\;\delta
_RA_a^{*\mu }=-\left( g_{ab}A^{b\mu }+\partial _\nu B_a^{\nu \mu }\right) , 
\end{equation}
\begin{equation}
\label{6}\delta _R\eta ^{*a\mu }=\varepsilon ^{\mu \nu \lambda \rho
}\partial _\nu B_{\lambda \rho }^{*a}, 
\end{equation}
\begin{equation}
\label{7}\delta _RC^{*a}=\partial _\mu \eta ^{*a\mu }. 
\end{equation}
The introduction of the antifields $C^{*a}$ is implied by the necessity to
`kill' the non trivial antighost number two co-cycles $\mu ^a=\partial _\mu
\eta ^{*a\mu }$ in the homology of $\delta _R$. The longitudinal complex
contains the pure ghost number one fermionic ghosts $\eta _{a\mu }$ and the
pure ghost number two bosonic ghosts for ghosts $C_a$. The definitions of $%
\gamma _R$ read as 
\begin{equation}
\label{7x}\gamma _RA_\mu ^a=0,\;\gamma _RB_a^{\mu \nu }=\varepsilon ^{\mu
\nu \lambda \rho }\partial _\lambda \eta _{\rho a},\;\gamma _R\eta _{a\mu
}=\partial _\mu C_a,\;\gamma _RC_a=0. 
\end{equation}
Extending $\delta _R$ on the ghosts through $\delta _R\eta _{a\mu }=0$, $%
\delta _RC_a=0$, and $\gamma _R$ on the antifields by $\gamma _RA_a^{*\mu
}=0 $, $\gamma _RB_{\mu \nu }^{*a}=0$, $\gamma _R\eta ^{*a\mu }=0$, $\gamma
_RC^{*a}=0$, we find that $s_R^2=0$, $H^0\left( s_R\right) =\left\{ {\rm %
physical\;observables}\right\} $, where $H^0\left( s_R\right) $ represents
the zeroth order cohomological group of $s_R$.

The main idea underlying our construction is to redefine the antifields $%
\eta ^{*a\mu }$ in such a way that the new co-cycles of the type $\mu ^a$
identically vanish. If this is done, then the antifields $C^{*a}$ are
useless as there are no longer any non trivial co-cycles at antighost number
two. In this way, we infer an irreducible Koszul-Tate complex, which further
leads to a longitudinal complex that contains no more the ghosts for ghosts $%
C_a$. Accordingly our idea, we redefine the antifields $\eta ^{*a\mu }$ like 
\begin{equation}
\label{9}\eta ^{*a\mu }\rightarrow \eta ^{\prime *a\mu }=M_{\;\;b\nu }^{a\mu
}\eta ^{*b\nu }, 
\end{equation}
where $M_{\;\;b\nu }^{a\mu }$ are taken to satisfy the conditions 
\begin{equation}
\label{10}\partial _\mu M_{\;\;b\nu }^{a\mu }=0, 
\end{equation}
\begin{equation}
\label{11}M_{\;\;b\nu }^{a\mu }\varepsilon ^{\nu \sigma \lambda \rho
}\partial _\sigma B_{\lambda \rho }^{*b}=\varepsilon ^{\mu \sigma \lambda
\rho }\partial _\sigma B_{\lambda \rho }^{*a}. 
\end{equation}
With the help of (\ref{6}), (\ref{9}) and (\ref{11}) we find that 
\begin{equation}
\label{12}\delta \eta ^{\prime *a\mu }=\varepsilon ^{\mu \sigma \lambda \rho
}\partial _\sigma B_{\lambda \rho }^{*a}. 
\end{equation}
The last equations do not further imply non trivial co-cycles because the
new co-cycles of the type $\mu ^a$ identically vanish via (\ref{10}), hence
we passed to an irreducible situation. In (\ref{12}) we employed the
notation $\delta $ instead of $\delta _R$ in order to emphasize that the
Koszul-Tate complex becomes irreducible. The solution to (\ref{9}--\ref{10})
is expressed by 
\begin{equation}
\label{13}M_{\;\;b\nu }^{a\mu }=\delta _{\;\;b}^a\left( \delta _{\;\;\nu
}^\mu -\frac{\partial ^\mu \partial _\nu }{\Box }\right) , 
\end{equation}
where $\Box =\partial _\lambda \partial ^\lambda $. Substituting (\ref{13})
in (\ref{12}), we get 
\begin{equation}
\label{14}\delta \left( \eta ^{*a\mu }-\frac{\partial ^\mu \partial _\nu }{%
\Box }\eta ^{*a\nu }\right) =\varepsilon ^{\mu \sigma \lambda \rho }\partial
_\sigma B_{\lambda \rho }^{*a}. 
\end{equation}
At this stage we introduce some scalar fields $\varphi _a$ whose antifields $%
\varphi ^{*a}$ are demanded to be the non vanishing solutions to the
equations 
\begin{equation}
\label{15}-\Box \varphi ^{*a}=\delta \left( \partial _\mu \eta ^{*a\mu
}\right) . 
\end{equation}
The non vanishing solutions $\varphi ^{*a}$ enforce the irreducibility as (%
\ref{15}) possess non vanishing solutions if and only if $\delta \left(
\partial _\mu \eta ^{*a\mu }\right) \neq 0$, therefore if and only if $\mu
^a $ are no longer co-cycles. Using (\ref{14}--\ref{15}) we find that 
\begin{equation}
\label{16}\delta \eta ^{*a\mu }=\varepsilon ^{\mu \sigma \lambda \rho
}\partial _\sigma B_{\lambda \rho }^{*a}-\partial ^\mu \varphi ^{*a}. 
\end{equation}
In order to preserve the nilpotency of $\delta $ we set 
\begin{equation}
\label{17}\delta \varphi ^{*a}=0. 
\end{equation}
If we maintain the actions of $\delta $ like in the reducible case 
\begin{equation}
\label{18}\delta B_a^{\mu \nu }=0,\;\delta A_\mu ^a=0, 
\end{equation}
\begin{equation}
\label{19}\delta B_{\mu \nu }^{*a}=\frac 12F_{\mu \nu }^a,\;\delta A_a^{*\mu
}=-\left( g_{ab}A^{b\mu }+\partial _\nu B_a^{\nu \mu }\right) , 
\end{equation}
and define 
\begin{equation}
\label{20}\delta \varphi _a=0, 
\end{equation}
then the formulas (\ref{16}--\ref{20}) describe an irreducible Koszul-Tate
complex. We remark that the irreducibility was gained by introducing the
supplementary fields $\varphi _a$ and their antifields $\varphi ^{*a}$ in
the theory. From (\ref{16}--\ref{20}) we can derive the Lagrangian action
and the gauge transformations of the irreducible theory. If we denote by $%
\tilde S_0^L\left[ A_\mu ^a,B_a^{\mu \nu },\varphi _a\right] $ the
Lagrangian action of the irreducible model, then by means of the general
relations $\delta \varphi ^{*a}=-\delta \tilde S_0^L/\delta \varphi _a$ and (%
\ref{17}) we obtain that 
\begin{equation}
\label{22}\tilde S_0^L\left[ A_\mu ^a,B_a^{\mu \nu },\varphi _a\right]
=S_0^L\left[ A_\mu ^a,B_a^{\mu \nu }\right] , 
\end{equation}
such that the dependence on $\varphi _a$ is trivial. On the other hand, with
the help of the original gauge transformations and (\ref{16}), it results
that the gauge transformations of the irreducible system are expressed by 
\begin{equation}
\label{23}\delta _\epsilon B_a^{\mu \nu }=\varepsilon ^{\mu \nu \lambda \rho
}\partial _\lambda \epsilon _{\rho a},\;\delta _\epsilon A_\mu ^a=0,\;\delta
_\epsilon \varphi _a=\partial ^\mu \epsilon _{a\mu }. 
\end{equation}
In this manner, we derived an irreducible theory based on action (\ref{22})
and the irreducible gauge transformations (\ref{23}) associated with the
abelian Freedman-Townsend model. From (\ref{22}) we notice that the newly
added fields $\varphi _a$ are not involved with the Lagrangian action of the
irreducible theory, hence they are purely gauge. As a consequence, the
physical observables (gauge invariant functions) of the irreducible model do
not depend on the $\varphi _a$'s and, in addition, are invariant under the
gauge transformations $\delta _\epsilon B_a^{\mu \nu }=\varepsilon ^{\mu \nu
\lambda \rho }\partial _\lambda \epsilon _{\rho a},\;\delta _\epsilon A_\mu
^a=0$, so they coincide with the physical observables of the original
redundant theory. The construction of the irreducible longitudinal
differential along the gauge orbits, $\gamma $ is realized via the
definitions 
\begin{equation}
\label{24}\gamma B_a^{\mu \nu }=\varepsilon ^{\mu \nu \lambda \rho }\partial
_\lambda \eta _{\rho a},\;\gamma A_a^\mu =0,\;\gamma \varphi _a=\partial
^\mu \eta _{a\mu },\;\gamma \eta _{a\mu }=0, 
\end{equation}
such that $\gamma $ is nilpotent, $\gamma ^2=0$, without introducing the
ghosts for ghosts. If we extend $\delta $ to the ghosts through $\delta \eta
_{a\mu }=0$ and $\gamma $ to the antifields by $\gamma B_{\mu \nu }^{*a}=0$, 
$\gamma A_a^{*\mu }=0$, $\gamma \varphi ^{*a}=0$, $\gamma \eta ^{*a\mu }=0$,
then the homological perturbation theory \cite{h1}--\cite{h3} guarantees the
existence of the irreducible BRST symmetry $s_I=\delta +\gamma $ that is
nilpotent, $s_I^2=0$, and satisfies the property $H^0\left( s_I\right)
=\left\{ {\rm physical\;observables}\right\} $, where `physical observables'
are referring to the irreducible system. As we previously mentioned, the
physical observables corresponding to the reducible and irreducible
formulations coincide, which leads to $H^0\left( s_R\right) =H^0\left(
s_I\right) $, and moreover, the two Lagrangian BRST symmetries are nilpotent 
$s_R^2=0=s_I^2$. By virtue of the last two relations we conclude that the
two symmetries are equivalent from the BRST point of view, i.e., from the
point of view of the basic equations underlying the antifield-BRST
formalism. In consequence, we can replace the reducible Lagrangian BRST
symmetry with the irreducible one in the case of the model under study.

With the above conclusion at hand, we pass to the deformation procedure of
the irreducible version in the context of the antifield formalism \cite{def}%
. A consistent deformation of the free action $S_0^L\left[ A_\mu ^a,B_a^{\mu
\nu }\right] $ and of its gauge invariances defines a deformation of the
corresponding solution to the master equation that preserves both the master
equation and the field/antifield spectra. So, if $S_0^L\left[ A_\mu
^a,B_a^{\mu \nu }\right] +g\int d^4x\alpha _0+O\left( g^2\right) $ stands
for a consistent deformation of the free action, with deformed gauge
transformations $\bar \delta _\epsilon B_a^{\mu \nu }=\varepsilon ^{\mu \nu
\lambda \rho }\partial _\lambda \epsilon _{\rho a}+g\beta _a^{\mu \nu
}+O\left( g^2\right) $, $\bar \delta _\epsilon \varphi _a=\partial ^\mu
\epsilon _{a\mu }+g\beta _a+O\left( g^2\right) $, then the deformed solution
to the master equation 
\begin{equation}
\label{31}\bar S=S+g\int d^4x\alpha +O\left( g^2\right) , 
\end{equation}
satisfies $\left( \bar S,\bar S\right) =0$, where 
\begin{equation}
\label{30}S=S_0^L\left[ A_\mu ^a,B_a^{\mu \nu }\right] +\int d^4x\left(
\varepsilon ^{\mu \nu \lambda \rho }B_{\mu \nu }^{*a}\partial _\lambda \eta
_{\rho a}+\varphi ^{*a}\partial ^\mu \eta _{a\mu }\right) , 
\end{equation}
and $\alpha =\alpha _0+B_{\mu \nu }^{*a}\bar \beta _a^{\mu \nu }+\varphi
^{*a}\bar \beta _a+`{\rm more}$'. Here, `${\rm more}$' stands for terms 'of
antighost number greater than one. The master equation $\left( \bar S,\bar
S\right) =0$ holds to order $g$ if and only if 
\begin{equation}
\label{30p}s_I\alpha =\partial _\mu j^\mu , 
\end{equation}
for some local $j^\mu $. This means that the non trivial first-order
consistent interactions belong to $H^0\left( s_I|d\right) $, where $d$ is
the exterior space-time derivative. In the case where $\alpha $ is a
coboundary modulo $d$ ($\alpha =s_I\rho +\partial _\mu b^\mu $), then the
deformation is trivial (it can be eliminated by a redefinition of the
fields). In order to investigate the solution to (\ref{30p}) we develop $%
\alpha $ accordingly the antighost number 
\begin{equation}
\label{31p}\alpha =\alpha _0+\alpha _1+\ldots ,\;antigh\left( \alpha
_k\right) =k, 
\end{equation}
where the last term from the sum can be assumed to be annihilated by $\gamma 
$. Because the free theory is irreducible, we can assume that $\alpha $
stops at antighost number one, i.e., $\alpha =\alpha _0+\alpha _1$, with $%
\alpha _1=\alpha ^{a\mu }\eta _{a\mu }$, where $\alpha ^{a\mu }$ pertains to 
$H_1\left( \delta |d\right) $, hence is a solution of the equation $\delta
\alpha ^{a\mu }+\partial _\rho \lambda ^{a\rho \mu }=0$. Like in the
reducible case \cite{23}, $H_2\left( \delta |d\right) $ does not vanish, but
the term $\alpha _2$ can be shown to vanish. Indeed, on the one hand $\alpha
_2$ is of the form $\alpha _2=\alpha ^{ab\mu \nu }\eta _{a\mu }\eta _{b\nu }$%
, where $\alpha ^{ab\mu \nu }$ belongs to $H_2\left( \delta |d\right) $. On
the other hand, the most general element in $H_2\left( \delta |d\right) $
reads as 
\begin{equation}
\label{31x}\alpha ^a=C_{\;\;bc}^a\left( \eta ^{*b\mu }A_\mu ^c+\frac
12\varepsilon ^{\mu \nu \rho \sigma }B_{\mu \nu }^{*b}B_{\rho \sigma
}^{*c}+g^{cd}\varphi ^{*b}\partial _\mu A_d^{*\mu }\right) , 
\end{equation}
with $g^{cd}$ the inverse of $g_{cd}$, which further gives that $\alpha
^{ab\mu \nu }=\alpha ^ah^{b\mu \nu }$, where $h^{b\mu \nu }$ are some
constants. By Lorentz covariance $\alpha ^{ab\mu \nu }$ must vanish,
therefore $\alpha _2$ also vanishes. Let us investigate now the term $\alpha
_1$. The general form of an object from $H_1\left( \delta |d\right) $ that
is annihilated by $\gamma $ reads as 
\begin{equation}
\label{32}\alpha ^{a\mu }=C_{\;\;bc}^a\left( \varphi ^{*b}f^{c\mu }\left(
A\right) +\varepsilon ^{\rho \nu \lambda \mu }B_{\rho \nu }^{*b}A_\lambda
^c\right) , 
\end{equation}
where $f^{c\mu }\left( A\right) $ is a function of $A_\mu ^a$ and $%
C_{\;\;bc}^a$ are some constants, with $C_{\;\;bc}^a=-C_{\;\;cb}^a$. It is
simple to see that $\delta \alpha ^{a\mu }=\partial _\rho \left( \frac
12C_{\;\;bc}^a\varepsilon ^{\rho \nu \lambda \mu }A_\nu ^bA_\lambda
^c\right) $, so $\alpha ^{a\mu }$ is in $H_1\left( \delta |d\right) $. On
the other hand, we obtain 
\begin{equation}
\label{34}\delta \alpha _1+\gamma \left( -\frac 12C_{\;\;bc}^aB_a^{\mu \nu
}A_\mu ^bA_\nu ^c\right) =\partial _\mu \left( -\frac
12C_{\;\;bc}^a\varepsilon ^{\mu \nu \lambda \rho }A_\nu ^bA_\lambda ^c\eta
_{a\rho }\right) . 
\end{equation}
If we compare the last equation with (\ref{30p}) at antighost number zero
(i.e., with the equation $\delta \alpha _1+\gamma \alpha _0=\partial _\mu
n^\mu $), it follows that 
\begin{equation}
\label{35}\alpha _0=-\frac 12C_{\;\;bc}^aB_a^{\mu \nu }A_\mu ^bA_\nu ^c. 
\end{equation}
Thus, the deformed solution to order $g$ reads as 
\begin{eqnarray}\label{36}
& &\bar S=S+g\int d^4x\left( -\frac 12C_{\;\;bc}^aB_a^{\mu \nu }A_\mu ^bA_\nu
^c+\right. \nonumber \\
& &\left. C_{\;\;bc}^a\left( \varphi ^{*b}f^{c\mu }\left( A\right)
+\varepsilon ^{\rho \nu \lambda \mu }B_{\rho \nu }^{*b}A_\lambda ^c\right)
\eta _{a\mu }\right) . 
\end{eqnarray}
If we compute the antibracket $\left( \bar S,\bar S\right) $ we obtain 
\begin{equation}
\label{38}\left( \bar S,\bar S\right) =\frac 13g^2C_{\;\;\left[ bc\right.
}^eC_{\;\;\left. d\right] e}^a\varepsilon ^{\mu \nu \lambda \rho }\int
d^4xA_\mu ^bA_\nu ^cA_\lambda ^d\eta _{a\rho }\equiv g^2\int d^4xu, 
\end{equation}
where $\left[ bcd\right] $ expresses the antisymmetry with respect to the
indices between brackets. If we denote the term in $g^2$ from (\ref{31}) by $%
g^2\int d^4xb$, then the interaction is consistent to order $g^2$ if and
only if $u=-s_Ib+\partial _\mu k^\mu $ \cite{def}. However, from (\ref{38})
we see that $u$ cannot be of that form, and so it must vanish. This means
that the constants $C_{\;\;bc}^a$ must fulfill the Jacobi identity 
\begin{equation}
\label{39}C_{\;\;\left[ bc\right. }^eC_{\;\;\left. d\right] e}^a=0, 
\end{equation}
hence must define the structure constants of a Lie algebra. In this
situation (\ref{38}) vanishes, so $\bar S$ (which is only of order $g$) is a
solution of the master equation without adding higher order terms in $g$
(the vanishing of $u$ implies that all the higher order terms vanish).

The terms from $\bar S$ that do not involve the antifields, $S_0^L\left[
A_\mu ^a,B_a^{\mu \nu }\right] -\frac 12C_{\;\;bc}^ag\int d^4xB_a^{\mu \nu
}A_\mu ^bA_\nu ^c$, give nothing but the well-known action of the
non-abelian Freedman-Townsend model, while $\bar S$ itself represents the
corresponding solution to the master equation deriving from our irreducible
BRST approach to this model. The terms from (\ref{36}) that are linear in
the antifields show that the deformed gauge transformations read as $\bar
\delta _\epsilon B_a^{\mu \nu }=\varepsilon ^{\mu \nu \lambda \rho }\left(
D_\lambda \right) _{\;\;a}^b\epsilon _{\rho b}$, $\bar \delta _\epsilon
A_\mu ^a=0$, $\bar \delta _\epsilon \varphi _a=\partial ^\mu \epsilon _{a\mu
}+gC_{\;\;ab}^cf^{b\mu }\left( A\right) \epsilon _{c\mu }$, such that the
gauge transformations for $B_a^{\mu \nu }$ and $A_\mu ^a$ take the familiar
form in the literature. In the above formulas, the covariant derivative is
defined by $\left( D_\lambda \right) _{\;\;a}^b=\delta _{\;\;a}^b\partial
_\lambda +gC_{\;\;ac}^bA_\lambda ^c$. In addition, we have derived a class
of gauge transformations for $\varphi _a$. We remark that the functions $%
f^{c\mu }\left( A\right) $ are still undetermined. They must be in such a
way that the deformed gauge transformations are irreducible. A choice that
preserves the irreducibility and, in the meantime, makes manifest the nice
structure represented by the covariant derivative is $f^{c\mu }\left(
A\right) =A^{c\mu }$, so $\bar \delta _\epsilon \varphi _a=\left( D^\mu
\right) _{\;\;a}^b\epsilon _{b\mu }$. The solution (\ref{36}) with $f^{c\mu
}\left( A\right) $ replaced by $A^{c\mu }$ differs from that obtained in the
literature by many authors \cite{19}--\cite{22} in the reducible framework.
The solution (\ref{36}) does not contain terms that are quadratic in the
antifields (like in the reducible situation), so the irreducible BRST
transformations $\bar s_IF=\left( F,\bar S\right) $ do not involve the
antifields, such that the gauge-fixed BRST symmetry does not depend on the
gauge-fixing fermion, by contrast with the reducible setting. In
consequence, our approach leads to a gauge-fixed BRST symmetry that is
off-shell nilpotent. Indeed, we have that $\bar s_IB_a^{\mu \nu
}=\varepsilon ^{\mu \nu \lambda \rho }\left( D_\lambda \right)
_{\;\;a}^b\eta _{b\rho }$, $\bar s_IA_\mu ^a=0$, $\bar s_I\varphi _a=\left(
D^\mu \right) _{\;\;a}^b\eta _{b\mu }$, $\bar s_I\eta _{a\mu }=0$. In the
meantime, the absence of the term quadratic in the antifields will
consequently imply the absence of the three-ghost coupling term in the
gauge-fixed action, such that the gauge-fixed action in the context of our
irreducible approach takes a simpler form. This completes our irreducible
procedure for deriving the Freedman-Townsend vertex.

To conclude with, in this letter we have exposed a cohomological approach to
the Freedman-Townsend model consisting in two basic steps, namely, the
construction of an irreducible BRST symmetry for the abelian version and the
subsequent deformation of the irreducible theory. The results arising in our 
{\it irreducible} procedure prove the uniqueness of the Freedman-Townsend
vertex in four dimensions (which has also been derived in \cite{23}, but
within the reducible background) and also lead to a deformed solution of the
master equation that has not previously been derived in the literature. In
this light, our irreducible approach represents an efficient alternative to
the reducible version exposed in \cite{23}.

\end{document}